\documentclass{appolb}
\usepackage{epsfig}

\begin{document}

\title{Molecular structures and clustering effects in reactions induced 
by light nuclei
\thanks{Presented at the First International African Symposium on Exotic Nuclei
IASEN2013 Ithemba LABS Cape Town, South Africa, December 2-6, 2013}
}

\author{C. Beck$^a$
\address{
$^a$D\'epartement de Recherches Subatomiques, Institut Pluridisciplinaire 
Hubert Curien, IN$_{2}$P$_{3}$-CNRS and Universit\'e de Strasbourg - 23, rue 
du Loess BP 28, F-67037 Strasbourg Cedex 2, France\\
E-mail: christian.beck@iphc.cnrs.fr\\}
}

\maketitle

\newpage

\begin{abstract}
A great deal of research work has been undertaken in 
$\alpha$-clustering study since the pioneering discovery of $^{12}$C+$^{12}$C 
molecular resonances half a century ago. Our knowledge on  
physics of nuclear molecules has increased considerably and nuclear 
clustering remains one of the most fruitful domains of nuclear physics,
facing some of the greatest challenges and opportunities in the years ahead. 
The occurrence of ``exotic" shapes in light $N$=$Z$ $\alpha$-like 
nuclei is investigated. Various approaches of the superdeformed and hyperdeformed 
bands associated with quasimolecular resonant structures are presented. 
Evolution of clustering from stability to the drip-lines is examined:
clustering aspects are, in particular, discussed for light exotic nuclei with
large neutron excess such as neutron-rich Oxygen isotopes with their complete 
spectrocopy.
\end{abstract}



\section{Introduction}

\begin{figure}
\includegraphics[scale= 0.55]{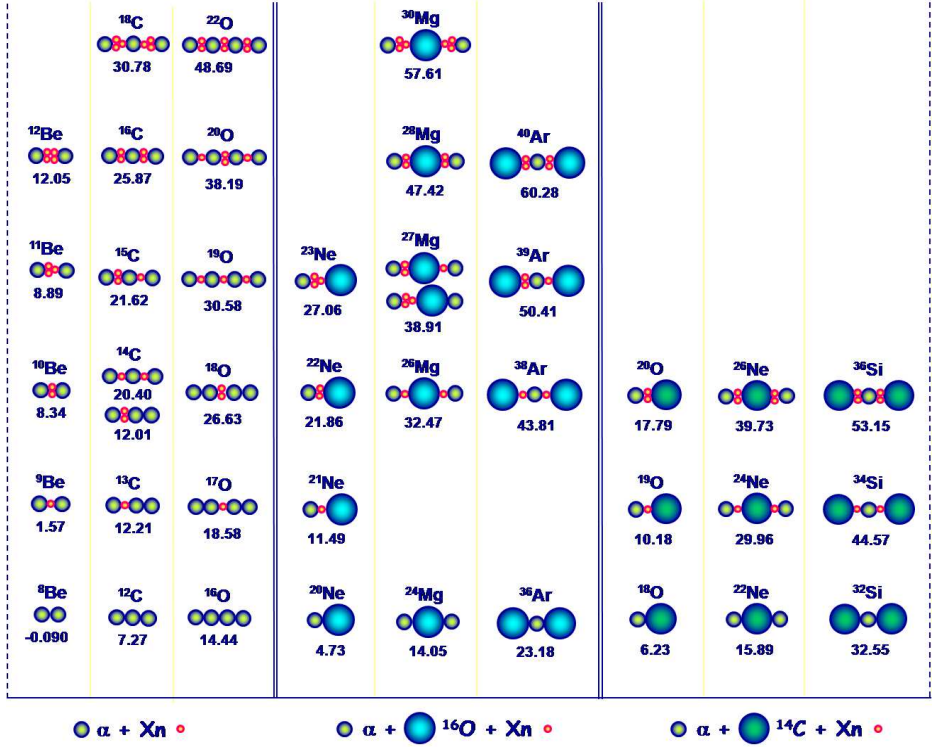}
\caption{\label{label}Schematic illustration of the structures of molecular
shape isomers in light neutron-rich isotopes of nuclei consisting
of $\alpha$-particles, $^{16}$O- and $^{14}$C-clusters plus some
covalently bound neutrons (Xn means X neutrons) \cite{Milin14}. The so called 
"Extended Ikeda-Diagram" \cite{Oertzen01} with $\alpha$-particles (left panel) and 
$^{16}$O-cores (middle panel) can be generalized to $^{14}$C-cluster cores 
(right panel. The lowest line of each configuration corresponds to parts
of the original Ikeda diagram \cite{Ikeda}. However, because of its deformation,
the $^{12}$C nucleus is not included, as it was earlier \cite{Ikeda}.
Threshold energies are given in MeV.}
\label{fig:1}
\end{figure}

One of the greatest challenges in nuclear science is the understanding of the
structure of light nuclei from both the experimental and theoretical perspectives.
Starting in the 1960s the search for resonant structures in the excitation functions for various 
combinations of light $\alpha$-cluster ($N$=$Z$) nuclei in the energy regime 
from the Coulomb barrier up to regions with excitation energies of $E_{x}$=20$-$50~MeV 
remains a subject of contemporary debate~\cite{Erb85,Greiner95,Beck94}. These 
resonances~\cite{Erb85} have been interpreted in terms of nuclear molecules~\cite{Greiner95}. 
The question of how quasimolecular resonances may reflect continuous transitions
from scattering states in the ion-ion potential to true cluster states in the 
compound systems was still unresolved in the 1990s \cite{Greiner95,Beck94}. In many 
cases, these resonant structures have been associated with strongly-deformed 
shapes and with $\alpha$-clustering phenomena \cite{Freer07,Horiuchi10}, predicted from the 
Nilsson-Strutinsky approach, the cranked $\alpha$-cluster model~\cite{Freer07}, or 
other mean-field calculations~\cite{Horiuchi10,Gupta10}. In light $\alpha$-like 
nuclei clustering is observed as a general phenomenon at high excitation energy 
close to the $\alpha$-decay thresholds \cite{Freer07,Oertzen06}. This exotic 
behavior has been perfectly illustrated by the famous ''Ikeda-diagram" for $N$=$Z$ 
nuclei in 1968 \cite{Ikeda}, which has been recently modified and extended by von Oertzen 
\cite{Oertzen01} for neutron-rich nuclei, as shown in the left panel of Fig.1.
Clustering is a general feature \cite{Milin14} not only observed in light
neutron-rich nuclei \cite{Kanada10}, but also in halo nuclei such as $^{11}$Li 
\cite{Ikeda10} or $^{14}$Be, for instance \cite{Nakamura12}. The problem of 
cluster formation has also been treated extensively for very heavy systems by 
R.G. Gupta \cite{Gupta10}, by Zagrebaev and W. Greiner \cite{Zagrebaev10} and
by C. Simenel \cite{Simenel14} where giant molecules and collinear ternary 
fission may co-exist \cite{Kamanin14}. Finally, signatures of $\alpha$ clustering have
also been discovered in light nuclei surviving from ultrarelativistic nuclear 
collisions \cite{Zarubin14,Broniowski14}.

\section{Alpha clustering, nuclear molecules and large deformations}
\label{sec:1}

The real link between superdeformation (SD), nuclear molecules and $\alpha$ 
clustering \cite{Horiuchi10,Beck04a,Beck04b,Beck04c,Cseh09} is of particular interest, 
since nuclear shapes with major-to-minor axis ratios of 2:1 have the typical 
ellipsoidal elongation for light nuclei i.e. with quadrupole deformation 
parameter $\beta_2$ $\approx$ 0.6. Furthermore, the structure of possible 
octupole-unstable 3:1 nuclear shapes - hyperdeformation (HD) with $\beta_2$ 
$\approx$ 1.0 - has also been discussed for actinide nuclei \cite{Cseh09} in 
terms of clustering phenomena. Typical examples for possible relationship 
between quasimolecular bands and extremely deformed (SD/HD) shapes have been 
widely discussed in the literature for $A = 20-60$ $\alpha$-conjugate 
$N$=$Z$ nuclei, such as $^{28}$Si 
\cite{Taniguchi09,Ichikawa11,Jenkins12,Jenkins14,Darai12}, $^{32}$S 
\cite{Horiuchi10,Kimura04,Lonnroth10,Chandana10,Ichikawa11}, 
$^{36}$Ar \cite{Cseh09,Beck08a,Svensson00,Sciani09,Beck09,Beck11,Beck13}, $^{40}$Ca 
\cite{Ideguchi01,Rousseau02,Taniguchi07,Norrby10,Benjamim13}, $^{44}$Ti 
\cite{Horiuchi10,Leary00,Fukada09}, $^{48}$Cr \cite{Salsac08} and $^{56}$Ni 
\cite{Nouicer99,Rudolph99,Beck01,Bhattacharya02}.

\begin{figure}
\includegraphics[scale=0.59]{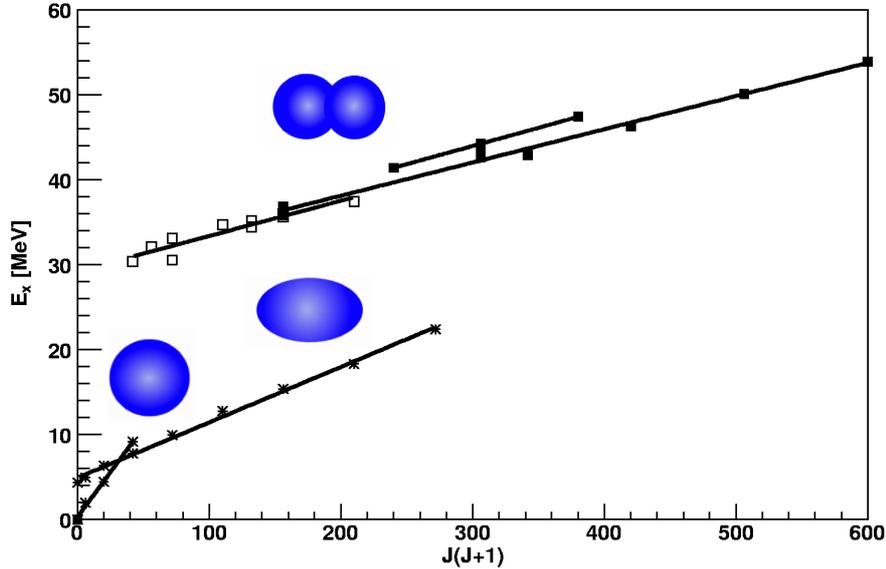}
\caption{\label{label}Rotational bands and deformed shapes in $^{36}$Ar. Excitation energies 
	of the ground state (spherical shape) and SD (ellipsoidal shape) bands~\cite{Svensson00}, respectively, and the energies of HD (dinuclear shape) band from 
	the quasimolecular resonances observed in the $^{12}$C+$^{24}$Mg 
	(open rectangles) \cite{Sciani09,Cindro79,Mermaz84,Pocanic85} and 
	$^{16}$O+$^{20}$Ne (full rectangles) \cite{Shimizu82,Gai84} reactions 
	are plotted as a function of J(J+1). This figure
	has been adapted from Refs.~\cite{Beck08a,Sciani09,Beck13}.}
\label{fig:2}
\end{figure}

In fact, highly deformed shapes and SD rotational bands have been 
discovered in several light $\alpha$-conjugate nuclei, such as $^{36}$Ar
and $^{40}$Ca by using $\gamma$-ray spectroscopy techniques 
\cite{Svensson00,Ideguchi01}. In particular, the extremely deformed rotational
bands in $^{36}$Ar \cite{Svensson00} (shown as crosses in Fig.~2) might be 
comparable in shape to the quasimolecular bands observed in both $^{12}$C+$^{24}$Mg 
\cite{Sciani09,Cindro79,Mermaz84,Pocanic85} (shown as open triangles in Fig.~2)
and $^{16}$O+$^{20}$Ne \cite{Shimizu82,Gai84} (shown as full rectangles) reactions. 
These resonances belong to a rotational band, with a moment of inertia close to 
that of a HD band provided by both the cranked $\alpha$-cluster model \cite{Freer07} 
and the Nilsson-Strutinsky \cite{Cseh09} calculations. The fact that similar
quasi-molecular states observed in the two reactions fall on the same rotational
band gives further support to our interpretation of the $^{36}$Ar composite system
resonances. An identical conclusion was reached for the $^{40}$Ca composite system
where SD bands have been discovered \cite{Svensson00}. Therefore, similar
investigations are underway for heavier $\alpha$-like composite systems such as
$^{44}$Ti 
\cite{Horiuchi10,Leary00,Fukada09}, $^{48}$Cr \cite{Salsac08} and $^{56}$Ni 
\cite{Nouicer99,Rudolph99,Beck01,Bhattacharya02}.

Ternary clusterizations in light $\alpha$-like composite systems are also 
predicted theoretically, but were not found experimentally in $^{36}$Ar so far 
\cite{Beck09}. On the other hand, ternary fission of $^{56}$Ni -- related to 
its HD shapes -- was identified from out-of-plane angular correlations 
measured in the $^{32}$S+$^{24}$Mg reaction with the Binary Reaction 
Spectrometer (BRS) at the {\sc Vivitron} Tandem facility of the IPHC, Strasbourg 
\cite{Oertzen08}. This finding \cite{Oertzen08} is not limited to light 
$N$=$Z$ compound nuclei, true ternary fission \cite{Zagrebaev10,Kamanin14,Pyatkov10}
can also occur for very heavy \cite{Kamanin14,Pyatkov10} and superheavy 
\cite{Zagrebaev10b} nuclei.

\section{Electromagnetic transitions as a probe of quasimolecular states
and clustering in light nuclei}

Clustering in light nuclei is traditionally explored through reaction studies,
but observation of electromagneetic transitions can be of high value in
establishing, for example, that highly-excited sates with candidate cluster
structure do indeed form rotational sequences.

There is a renewed interest in the spectroscopy of the $^{16}$O nucleus at high 
excitation energy \cite{Beck09}. Exclusive data were collected on $^{16}$O in the 
inverse kinematics reaction $^{24}$Mg$+^{12}$C studied at E$_{lab}$($^{24}$Mg) 
= 130 MeV with the BRS in coincidence with the {\sc Euroball IV} installed at 
the {\sc Vivitron} facility
\cite{Beck09}. From the $\alpha$-transfer reactions (both direct transfer
and deep-inelastic orbiting collisions \cite{Sanders99}), new information has been 
deduced on branching ratios of the decay of the 3$^{+}$ state of $^{16}$O at 
11.085~MeV $\pm$ 3 keV. The high-energy level scheme of $^{16}$O shown in
Ref.~\cite{Beck09} indicated that this state does not $\alpha$-decay because of its non-natural parity 
(in contrast to the two neighbouring 4$^{+}$ states at 10.36~MeV and 11.10~MeV), 
but it $\gamma$ decays to the 2$^{+}$ state at 6.92~MeV (54.6 $\pm$ 2 $\%$) and 
to the 3$^-$ state at 6.13~MeV (45.4\%). 
By considering all the four possible transition types of the decay of the 3$^{+}$ 
state (\textit{i.e.} E$1$ and M$2$ for the 3$^{+}$ $\rightarrow$ 3$^{-}$ transition and, 
M$1$ and E$2$ for the 3$^{+}$ $\rightarrow$ 2$^{+}$ transition), our 
calculations yield the conclusion that $\Gamma_{3^+}<0.23$~eV, a value 
fifty times lower than known previously, which is an important result for 
the well studied $^{16}$O nucleus \cite{Beck09}.
Clustering effects in the light neutron-rich 
oxygen isotopes $^{17,18,19,20}$O will also be discussed in Section 5.

\begin{figure}
\includegraphics[scale=0.93]{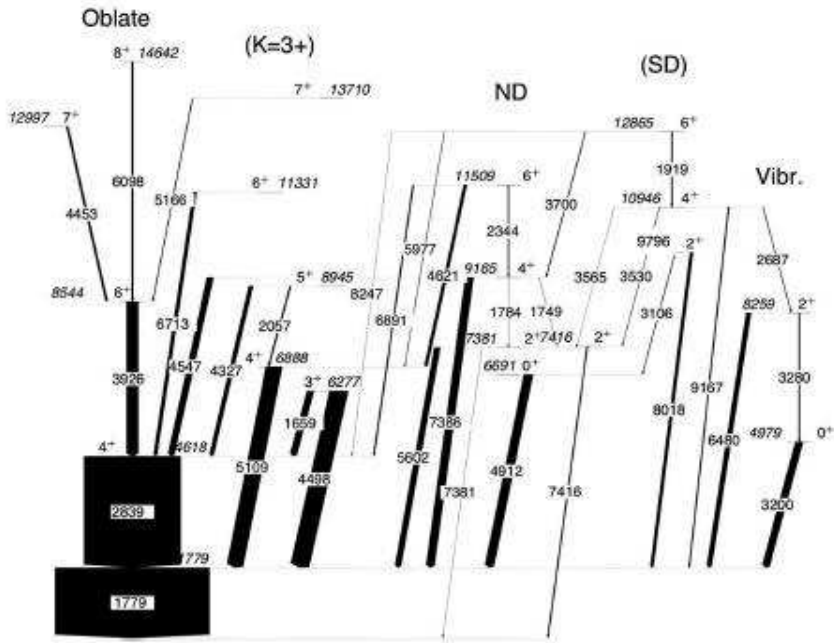}
\caption{\label{label} Subset of positive-parity levels in $^{28}$Si derived
from the analysis of a Gammasphere experiment. Excited states and transitions
energies are labeled with their energy in keV, while the width of the arrows
corresponds to the relatve intensity of the observed transitions. The
different structures are labeled according to previous assignements as oblate,
prolate (ND), vibrational and with different K values. This figure has adapted
from Ref.~\cite{Jenkins12}}
\label{fig:3}
\end{figure}

$\alpha$ clustering plays an important role in the description of the ground
state and excited states of light nuclei in the $p$ shell. For heavier nuclei,
in the $sd$-shell, cluster configurations may be based on heavier substructures
like $^{12}$C, $^{14}$C and $^{16}$O as shown by the ''Extended Ikeda-diagram"
proposed in Fig.~1. This was already well discussed to appear in 
$^{24}$Mg($^{12}$C-$^{12}$C) and $^{28}$Si($^{12}$C-$^{16}$O) both theoretically
and experimentally. The case of the mid-$sd$-shell nucleus $^{28}$Si is of
particular interest as it shows the coexistence of deformed and cluster states
at rather low energies \cite{Jenkins12}. Its ground state is oblate, with a 
partial $\alpha$-$^{24}$Mg structure, two prolate normal deformed bands are found, one
built on the ${0}^{+}_{2}$ state at 4.98 MeV and on the ${0}^{+}_{3}$ state 
at 6.69 MeV. The SD band candidate with a pronounced $\alpha$-$^{24}$Mg structure
is suggested \cite{Jenkins12}. In this band, the 2$^+$ (9.8 MeV), 4$^+$ and 6$^+$
members are well identified as can be clearly observed in Fig.~3.

In the following we will briefly discuss a resonant cluster band which is
predicted to start close to the Coulomb barrier of the $^{12}$C+$^{16}$O collision,
i.e. around 25 MeV excitation energy in $^{28}$Si \cite{Ichikawa11}. We have
studied the $^{12}$C($^{16}$O,$\gamma$)$^{28}$Si radiative capture reaction at
five resonant energies around the Coulomb barrier by using the zero degree
DRAGON spectrometer installed at Triumf, Vancouver \cite{Lebhertz12,Goasduff14}.
Details about the setup, that has been optimized for the 
$^{12}$C($^{12}$C,$\gamma$)$^{24}$Mg radiative capture reaction in our of previous
DRAGON experiments, can be found in Ref.~\cite{Jenkins07}. 
The $^{12}$C($^{16}$O,$\gamma$)$^{28}$Si data clearly show \cite{Lebhertz12,Goasduff14}
the direct feeding of the prolate 4$^{+}_{3}$ state at 9.16 MeV and the octupole 
deformed 3$^-$ at 6.88 MeV.
This state is the band head of an octupole band which mainly decays to the 
$^{28}$Si oblate ground state with a strong E$_3$ transition. Our results are 
very similar to what has been measured for the $^{12}$C+$^{12}$C radiative capture 
reaction above the Coulomb barrier in the first DRAGON experiment
\cite{Jenkins07} where the enhanced feeding of the $^{24}$Mg prolate band has 
been measured for a 4$^+$-2$^+$ resonance at E$_{c.m.}$ = 8.0 MeV near the 
Coulomb barrier.
At the lowest energy of $^{12}$C+$^{16}$O radiative capture reaction, an enhanced 
feeding from the resonance J$^{\pi}$ = 2$^+$ and 1$^+$ T=1 states around 11 MeV 
is observed in $^{28}$Si. Again this is consistent with $^{12}$C+$^{12}$O radiative capture 
reaction data where J$^{\pi}$ = 2$^+$ has been assigned to the entrance resonance
and an enhanced decay has been measured via intermediate 1$^+$ T=1 states around 
11 MeV in $^{24}$Mg. A definitive scenario for the decay of the resonances at these
low bombarding energies in both systems will come from the measurement of the
$\gamma$ decay spectra with a $\gamma$-spectrometer with better resolution than BGO
but still rather good efficiency such as LaBr$_3$ crystals.

\begin{figure}
\includegraphics[scale=0.57]{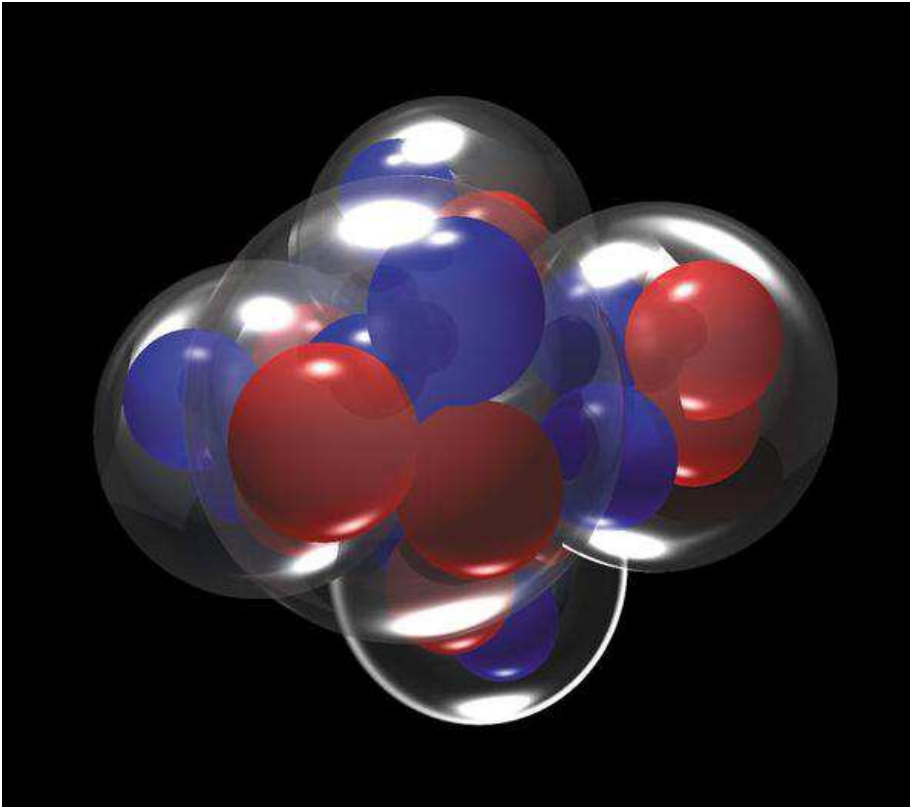}
\caption{\label{label} Schematic illustration of alpha clustering in $^{20}$Ne.}
\label{fig:4}
\end{figure}

\begin{figure}
\includegraphics[scale=0.42]{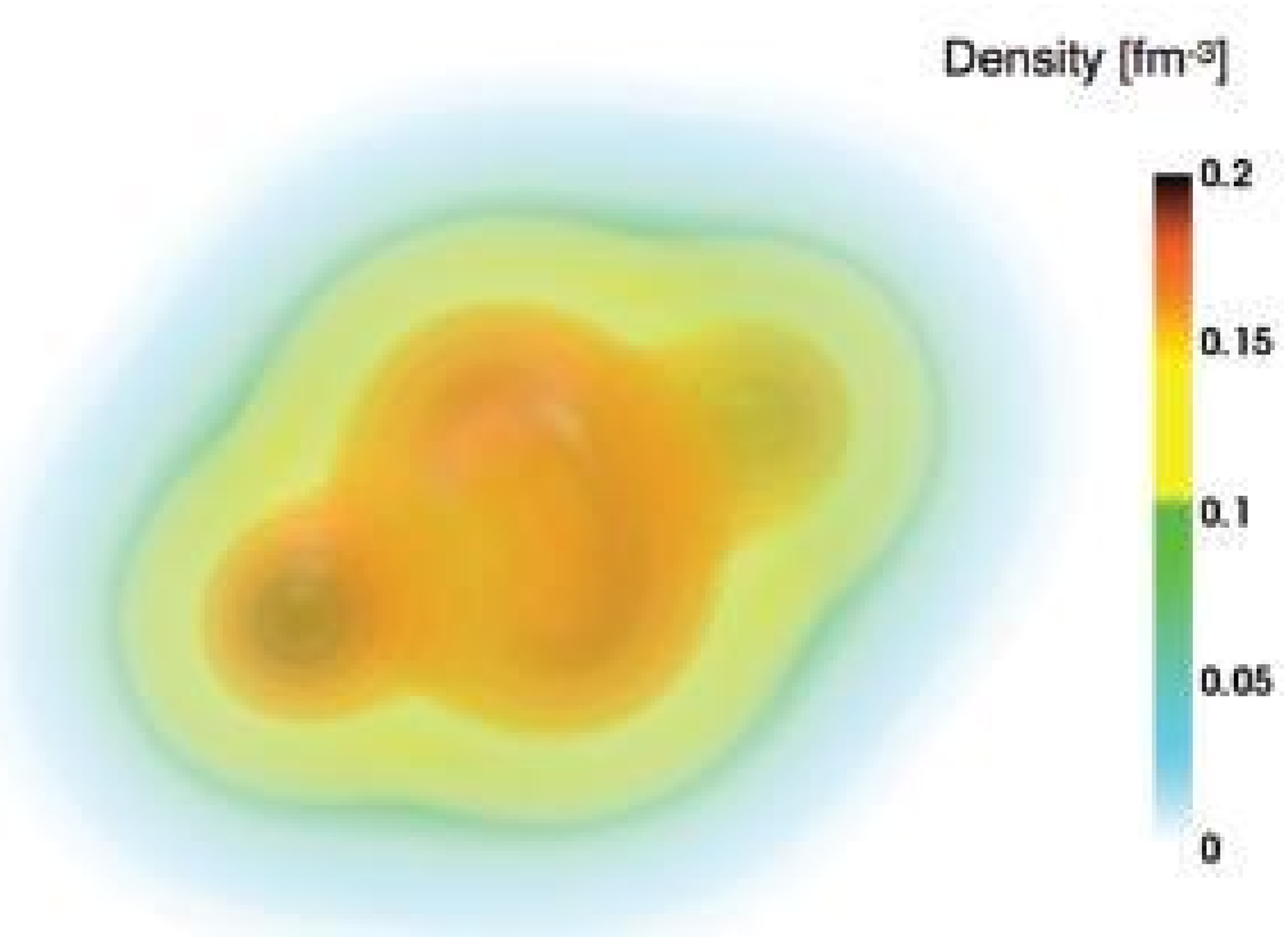}
\caption{\label{label} Self-consistent ground-state denisties of $^{20}$Ne 
as calculated with EDF. Densities (in units of fm$^{-3}$) are plotted in the
intrinsic frame of reference that coincides with the principal axes of the
nucleus. This figure has been adapted from Refs.~\cite{Ebran12}.}
\label{fig:5}
\end{figure}

\section{Condensation of $\alpha$ clusters in light nuclei}

In principle the nucleus is a quasi-homogeneus collection of protons and
neutrons, which adopts a spherical configuration i.e. a spherical droplet
of nuclear matter. For light nuclei the nucleons are capable to arrange
themselves into clusters of a bosonic character. The very stable
$\alpha$-particle is the most favorable light nucleus for quarteting -
$\alpha$ clustering - to occur in dense nuclear matter. These cluster 
structures have indeed a crucial role in the synthesis of elements in stars.
The so called ''Hoyle" state~\cite{Hoyle54}, the main 
portal through which $^{12}$C is created in nucleosynthesis with a 
pronounced three-$\alpha$-cluster structure, is the best exemple of
$\alpha$ clustering in light nuclei.

In $\alpha$ clustering a geometric picture can be proposed in the framework 
of point group symmetries \cite{Broniowski14}. For instance, in $^{8}$Be the 
two $\alpha$ clusters are separated by as much as $\approx$ 2fm, $^{12}$C 
exhibits a triangle arrangement of the three $\alpha$ particles $\approx$ 3fm 
apart, $^{16}$O forms a tetrahedron, etc. A very schematic picture of the 
$^{20}$Ne nucleus as an arrangement of five $\alpha$ particles is displayed
in Fig.~4 to illustrate the enhancement of the symmetries of the $\alpha$ 
clustering.

In the study of the Bose-Einstein Condensation (BEC) the
$\alpha$-particle states were first described for $^{12}$C and $^{16}$O
\cite{Tohsaki01,Suhara14} and later on generalized to heavier light $N$=$Z$ nuclei 
\cite{Oertzen10a,Yamada12,Ebran12,Girod13}. The structure of the ``Hoyle"
state and the properties of its assumed rotational band have been studied
very carefully from measurements of the $^{12}$C($\gamma$,3$\alpha$) reaction
performed at the HIGS facility, TUNL~\cite{Zimmermann13}. 

At present, the search for an experimental signature of BEC in $^{16}$O is 
of highest priority. A state with the structure of the ''Hoyle" state 
in $^{12}$C coupled to an $\alpha$ particle is predicted 
in $^{16}$O at about 15.1 MeV (the 0$^{+}_{6}$ state), the energy of which 
is $\approx$ 700 keV above the 4$\alpha$-particle breakup threshold 
\cite{Funaki08,Dufour13,Kanada14}: in other words, this 0$^{+}_{6}$ state 
might be a good candidate for the dilute 4$\alpha$ gas state.
However, any state in $^{16}$O equivalent to the ''Hoyle" 
state in $^{12}$C is most certainly going to decay by particle emission with 
very small, probably un-measurable, $\gamma$-decay branches, thus, very 
efficient particle-detection techniques will have to be used in the near 
future to search for them. 

BEC states are expected to decay by $\alpha$ emission to the ''Hoyle" state 
and could be found among the resonances in $\alpha$-particle inelastic 
scattering on $^{12}$C decaying to that state or could be observed in an 
$\alpha$-particle transfer channel leading to the $^{8}$Be--$^{8}$Be final state. 
The early attempt to excite these states by $\alpha$ inelastic 
scattering~\cite{Itoh04} was confirmed recently in Ref.~\cite{Itoh11,Freer12}.
Another possibility, that has not been yet explored, might be to perform Coulomb 
excitation measurements with intense $^{16}$O beams at intermediate energies.

Clustering of $^{20}$Ne has also been described within the density functional
theory~\cite{Ebran12} (EDF) as illustrated by Fig.~4 that displays axially and
reflection symmetric self-consistent equilibrium nucleon density distributions.
We note the well known quasimolecular $\alpha$-$^{12}$C-$\alpha$ structure
although clustering effects are less pronounced than the ones (schematically 
displayed in Fig.~3) predicted by Nilsson-Strutinsky calculations and 
even by mean-field calculations (including Hartree-Fock and/or 
Hartree-Fock-Bogoliubov calculations) \cite{Freer07,Horiuchi10,Gupta10,Girod13}.

The most recent work of Girod and Schuck \cite{Girod13} validates several
possible scenarios for the influence of clustering effects as a function of 
the neutron richness that will trigger more experimental works. We describe 
in the following Section recent experimental investigations on the Oxygen
isotopes chain.

\begin{figure}
\includegraphics[scale=0.62]{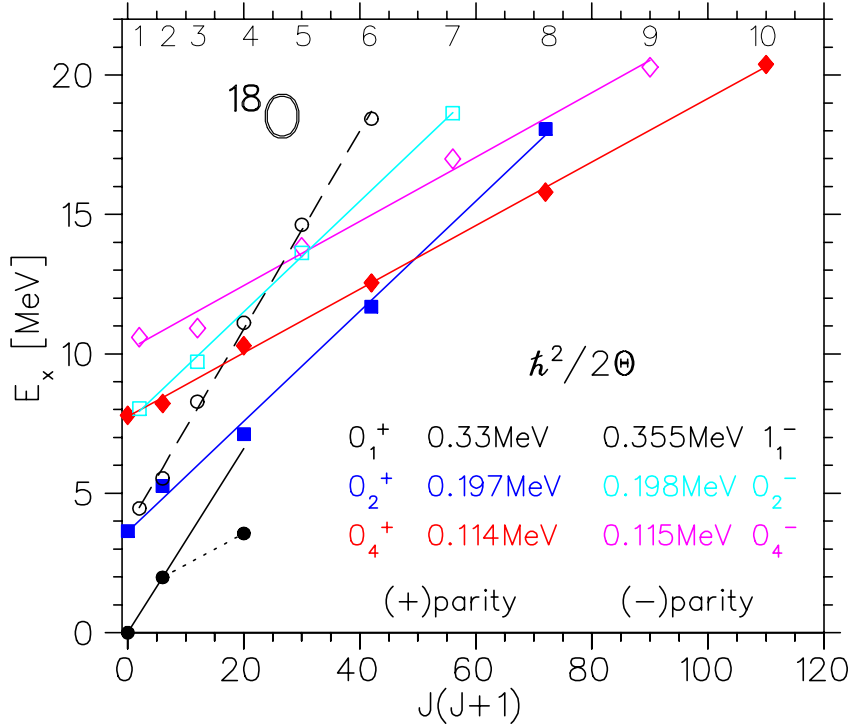}
\caption{\label{label}Overview of six rotational band structures observed in 
	$^{18}$O. Excitation energy systematics for the members of the rotational
	bands forming inversion doublets with K=0 are plotted as a function 
	of J(J+1). The curves are drawn to guide the eye for the slopes. The 
	indicated slope parameters contain information on
	the moments of inertia. Square symbols correspond to cluster bands,
	whereas diamonds symbols correspond to molecular bands. This figure
	is adapted from \cite{Oertzen10b}.}
\label{fig:6}
\end{figure}

\begin{figure}
\includegraphics[scale=0.62]{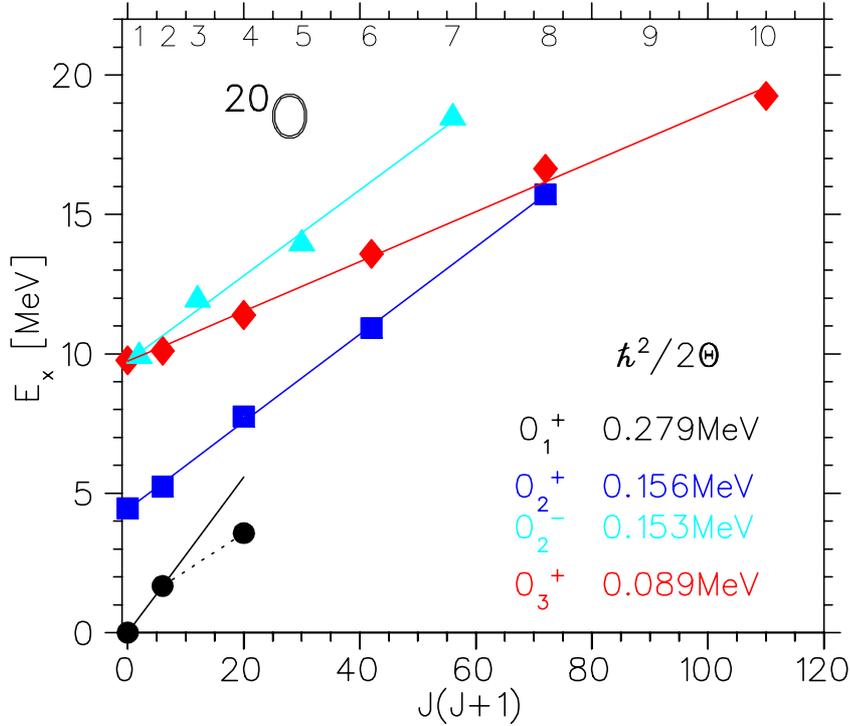}
\caption{\label{label}Overview of four rotational band structures observed in 
        $^{20}$O. Symbols are as in Fig.~6.This figure is adapted from \cite{Oertzen09}}
\label{fig:7} 
\end{figure}

\section{Clustering in light neutron-rich nuclei}
\label{sec:2}

As discussed previously, clustering is a general phenomenon observed also in 
nuclei with extra neutrons as it is presented in an ''Extended Ikeda-diagram" 
\cite{Ikeda} proposed by von Oertzen \cite{Oertzen01} (see the left panel of 
Fig.~1).
With additional neutrons, specific molecular structures 
appear with binding effects based on covalent molecular neutron orbitals. In 
these diagrams $\alpha$-clusters and $^{16}$O-clusters (as shown by the middle
panel of the diagram of Fig.~1) are the main ingredients. Actually, the $^{14}$C 
nucleus may play similar role in clusterization as the $^{16}$O one since it has similar  properties as a cluster: i) it has closed neutron p-shells, ii) first excited states are well above E$^{*}$ = 6 MeV, and iii) it has high binding energies for $\alpha$-particles.

A general picture of clustering and molecular configurations in light nuclei 
can be drawn from a detailed investigation of the light oxygen isotopes with
A $\geq$ 17. Here we will only present recent results on the even-even 
oxygen isotopes: $^{18}$O \cite{Oertzen10b} and $^{20}$O \cite{Oertzen09}. 
But very striking cluster states have also been found in odd-even oxygen 
isotopes such as: $^{17}$O \cite{Milin09} and $^{19}$O \cite{Oertzen11}. 

Fig.~6 gives an overview of all bands in $^{18}$O as a plot of excitation energies
as a function of J(J+1) together with their respective moments of inertia. In the
assignment of the bands both the dependence of excitation energies on J(J+1)
and the dependence of measured cross sections on 2J+1 \cite{Oertzen10b}
were considered. Slope parameters obtained in
a linear fit to the excitation energies \cite{Oertzen10b} indicate the moment
of inertia of the rotational bands given in Fig.~6. The intrinsic structure
of the cluster bands is reflection asymmetric, the parity projection gives an 
energy splitting between the partner bands.  
The assignment of the experimental molecular bands are supported by either
generator-coordinate-method \cite{Descouvemont} or Antisymmetrized Molecular
Dynamics (AMD) calculations \cite{Furutachi08}. 

We can compare the bands of $^{20}$O \cite{Oertzen09} shown in Fig.~7 with those of $^{18}$O
displayed in Fig.~6. The first doublet (K=0$^{\pm}_{2}$) has a slightly larger moment of
inertia (smaller slope parameter) in $^{20}$O, which is consistent
with its interpretation as $^{14}$C--$^{6}$He or $^{16}$C--$^{4}$He 
molecular structures (they start well below the thresholds of 16.8 MeV and 
12.32 MeV, respectively). The second band, for which the negative parity 
partner is yet to be determined, has a slope parameter slightly smaller
than in $^{18}$O. This is consistent with the study of the bands in 
$^{20}$O by Furutachi et al. \cite{Furutachi08}, which clearly establishes 
parity inversion doublets predicted by AMD calculations for the 
$^{14}$C--$^6$He cluster and $^{14}$C-2n-$\alpha$ molecular structures.
The corresponding moments of inertia given in Fig.~6 and Fig.~7 are 
strongly suggesting large deformations for the cluster structures. We may
conclude that the  reduction of the moments of inertia of the lowest
bands of $^{18,20}$O is consistent with the assumption that the strongly bound $^{14}$C 
nucleus having equivalent properties to $^{16}$O, has a similar role
as $^{16}$O in relevant, less neutron rich nuclei. Therefore, the Ikeda-diagram 
\cite{Ikeda} and the "extended Ikeda-diagram" consisting of $^{16}$O cluster
cores with covalently bound neutrons \cite{Oertzen01} must be further extended to 
include also the $^{14}$C cluster cores as illustrated in Fig.~1. 

\section{Summary, conclusions and outlook}

The link of $\alpha$-clustering, quasimolecular resonances, orbiting 
phenomena and extreme deformations (SD, HD, ...) has been discussed in this 
work. In particular, by using $\gamma$-ray spectroscopy of coincident binary 
fragments from 
either inelastic excitations and direct transfers in the $^{24}$Mg+$^{12}$C reaction. From a careful analysis of the $^{16}$O+$^{20}$Ne 
$\alpha$-transfer exit-channel (strongly populated by orbiting) new information 
has been deduced on branching ratios of the decay of the 3$^{+}$ state of $^{16}$O 
at 11.089~MeV. This result is encouraging for a complete spectroscopy 
of the $^{16}$O nucleus at high excitation energy. New results regarding cluster
and molecular states in neutron-rich oxygen isotopes in agreement with AMD 
predictions are presented. Consequently, the ''Extended Ikeda-diagram"
has been further modified for light neutron-rich nuclei by inclusion of the $^{14}$C 
cluster, similarly to the $^{16}$O one.  Of particular interest is the 
quest for the  4$\alpha$ states of $^{16}$O near the $^{8}$Be+$^{8}$Be and 
$^{12}$C+$\alpha$ decay thresholds, which correspond to the so-called ''Hoyle" 
state. 
The search for extremely elongated configurations (HD) in rapidly rotating medium-mass nuclei, which has 
been pursued by $\gamma$-ray spectroscopy measurements, will have to be 
performed in conjunction with charged-particle techniques in the near future 
since such states are most certainly  going to decay by particle emission 
(see \cite{Oertzen08,Papka12}).
Marked progress has been made in many traditional and novels subjects of nuclear
cluster physics. The developments in thse subjects show the importance of
clustering among the basic modes of motion of nuclear many-body systems.
All thess open questions will require precise coincidence measurements 
\cite{Papka12}
coupled with state-of-the-art theory.

\section{Acknowledments}

I would like to acknowledge Christian Caron (Springer) for initiating 
in 2008 the series of the three volumes of \emph{Lecture Notes in Physics} that
I have dedicated to "Clusters in Nuclei" between 2010 and 2014: the authors of
all the chapters (see Refs.
\cite{Horiuchi10,Gupta10,Milin14,Kanada10,Ikeda10,Nakamura12,Zagrebaev10,Simenel14,Kamanin14,Zarubin14,Jenkins14,Oertzen10a,Yamada12,Papka12})
are also warmy thanked for their fruitful collaboration during
the completion of this project.



\end{document}